\documentclass[11pt]{article}
\usepackage{graphicx}
\usepackage{amssymb}
\usepackage{color}
\usepackage{amsmath}
\usepackage{amsfonts}
\usepackage{amsthm}
\usepackage{hyperref}
\usepackage[mathscr]{eucal}
\usepackage{enumerate}
\usepackage{times}
\usepackage{authblk}

\usepackage{hyperref}
\usepackage{epigraph}
\usepackage{soul}

\newtheorem{Def}{Definition}[section]
\newtheorem{Thm}[Def]{Theorem}

\newcommand{\beq}{\begin{equation}}
\newcommand{\eeq}{\end{equation}}
\newcommand{\Proof}{\begin{proof}}
\newcommand{\QED}{\end{proof} \noindent}

\newcommand{\mm}{\hspace{-.08cm}\cdot \hspace{-.08cm}}

\newcommand{\Gammati}{\tilde{\Gamma}}

\newcommand{\A}{\mathcal{A}}
\newcommand{\Ati}{\tilde{\mathcal{A}}}
\newcommand{\Aop}{\mathcal{A}_\textbf{b}}

\title{Optimal Regularity and Uhlenbeck Compactness for General Relativity and Yang-Mills Theory}

\author[$\dagger$]{Moritz Reintjes}
\affil[$\dagger$]{Department of Mathematics\\ City University of Hong Kong\\ Kowloon \\ Hong Kong\\ moritzreintjes@gmail.com}

\author[$\star$]{Blake Temple}
\affil[$\star$]{Department of Mathematics\\ University of California\\ Davis, CA 95616\\ USA\\ temple@math.ucdavis.edu}

\date{24 June 2022}

\begin{document}
\maketitle

\begin{abstract} 
We announce the extension of {\it optimal regularity} and {\it Uhlenbeck compactness} to the general setting of connections on vector bundles with non-compact gauge groups over non-Riemannian manifolds, including the Lorent\-zian metric connections of General Relativity. Compactness is the essential tool of mathematical analysis for establishing validity of approximation schemes. Our proofs are based on the theory of the RT-equations for connections with $L^p$ curvature. Solutions of the RT-equations furnish coordinate and gauge transformations which give a non-optimal connection a gain of one derivative over its Riemann curvature, (i.e., to optimal regularity). The RT-equations are {\it elliptic} regardless of metric signature, and regularize singularities in solutions of the {\it hyperbolic} Einstein equations.  As an application, singularities at GR shock waves are removable, implying geodesic curves, locally inertial coordinates and the Newtonian limit all exist.   By the extra derivative we extend Uhlenbeck compactness from Uhlenbeck's setting of vector bundles with compact gauge groups over Riemannian manifolds, to the case of compact and non-compact gauge groups over non-Riemannian manifolds.  Our version of Uhlenbeck compactness can also be viewed as a ``geometric'' improvement of the Div-Curl Lemma, improving weak continuity of wedge products to {\it strong} convergence. 
\end{abstract}

{\quote\it``[Riemann...bound by Dirichlet...] would give acute, logical analyses of foundational questions, and would avoid long computations as much as possible.''} \hfill ---Felix~Klein

\section{Introduction}  

In his celebrated habilitation of 1854,  ``{\it On the hypotheses which lie at the foundations of geometry}'', Bernhard Riemann explained how to solve the longstanding problem proposed to him by Gauss, the problem of defining curvature in spaces of dimension larger than two.   His idea was to discover an essential mathematical construct formed from second derivatives of a metric, such that it transformed under coordinate change like a first derivative object--what came to be known as a {\it tensor}.   Thus began one of the great efforts of modern mathematics, to set Riemann's ideas upon a solid mathematical framework, and extend them to Physics.  Levi-Civita explained curvature in terms of ``parallel translation'' by the covariant derivative defined in terms of a {\it connection}, and the connection replaced the metric as the starting point of Riemann's theory of curvature. By this, Riemann's theory of curvature applies to any connection defined on the tangent bundle of an arbitrary differentiable manifold, (an {\it affine} connection), and extends naturally to connections on vector bundles over such manifolds. In the special case of metric connections, the connection determines the regularity of the metric.\footnote{Namely, first derivatives of the metric can be expressed in terms of the connection via Christoffel's formula, implying that the metric is precisely one derivative more regular than its connection in every coordinate system.}  In 1915 Albert Einstein introduced General Relativity (GR), and based his equations on Riemann's theory of curvature in Lorentzian geometry, where the metric connection describes the parallel translation of non-rotating frames. However, a fundamental consequence of the tensorial nature of Riemann's curvature is that it implies the existence of low regularity coordinate transformations which leave the regularity of the curvature unchanged, but lower the regularity of the connection to that of the curvature, thus making the connection ``singular'', (i.e., {\it non-optimal}).  This produces an ambiguity in the regularity of a connection, and hence of metrics as well.  When presented with a non-optimal connection (or metric), no regularizing coordinate transformation is given, and it has long been an open question whether all such singularities are removable by coordinate transformation, or whether classes of them exist which are ``geometric''.   This is fundamental at low regularity, for example, because one needs the connection to be one derivative above its curvature to conclude the strong convergence of approximation sequences;  and  the extra derivative is required to evolve solutions within the framework of nonlinear wave equations.  At the low regularity of GR shock waves, the singularities are so severe that they call into question the physical significance of the solutions themselves.

In this paper we summarize our recent resolution of the problem of {\it optimal regularity} for general connections on vector bundles with compact and non-compact gauge groups and with non-Riemannian affine part, including the Lorentzian metric connections of Relativity. The extra derivative implies Uhlenbeck compactness as a Corollary, which is the second main result announced in this paper. The proofs of these results for (non-Riemannian) affine connections are presented in \cite{ReintjesTemple_Uhl1}, and the extension to vector bundles is accomplished in \cite{ReintjesTemple_Uhl2}. The proofs are based on the RT-equations, a system of elliptic partial differential equations first introduced in \cite{ReintjesTemple_ell1} which determines the coordinate transformations to optimal regularity, c.f. \cite{ReintjesTemple_ell3}.

The RT-equations are a system of differential equations which are elliptic regardless of metric and metric signature.   By an existence theorem for the RT-equations (based on the $L^p$ theory of elliptic regularity), we establish optimal regularity for general connections, including both affine connections on arbitrary manifolds, and connections defined on vector bundles over such manifolds, including the Lorentzian manifolds of General Relativity.  Remarkably, the {\it elliptic} RT-equations determine the optimal regularity of solutions of the {\it hyperbolic} Einstein equations. For affine connections, we prove in \cite{ReintjesTemple_Uhl1} that any connection $\Gamma$ defined on the tangent bundle of an $n$-dimensional differentiable manifold, such that it satisfies the condition that its components $\Gamma\in L^{2p}$ and the components of its Riemann curvature tensor ${\rm Riem}(\Gamma)\in L^p$ in some coordinate system, $p>n/2$, can always be smoothed locally by a $W^{1,2p}$ coordinate transformation to optimal regularity $\Gamma\in W^{1,p}$; i.e.,  $\Gamma$ one order of differentiability above its curvature ${\rm Riem}(\Gamma)$.   In \cite{ReintjesTemple_Uhl2} we derive a vector bundle version of the RT-equations, and prove optimal regularity for connections on vector bundles over arbitrary base manifolds, in the same Sobolev spaces, allowing for both compact and non-compact gauge groups.  (See Theorem \ref{Thm_opt} below.)  In seminal work, Kazdan and DeTurck \cite{DeTurckKazdan} resolved the problem of optimal regularity for Riemannian (positive definite) metric geometries.  Our theory of the RT-equations extends this to Lorentzian geometry, and more generally to connections on tangent bundles and on vector bundles.  As a corollary, this resolves the long-standing open problem of the essential regularity of GR shock waves.  By the gain of one derivative, our theory also extends Uhlenbeck's celebrated compactness theorem from Uhlenbeck's setting of vector bundles with compact Lie groups over Riemannian manifolds \cite{Uhlenbeck,Wehrheim}, to general vector bundles with compact or non-compact Lie groups, over arbitrary manifolds.   Taken together, the theory of the RT-equations extends optimal regularity and Uhlenbeck compactness from Riemannian geometry, to the Lorentzian geometries of General Relativity and Yang-Mills Theory, the setting of Mathematical Physics.\footnote{Uhlenbeck's work on compactness for $L^p$ connections in Riemannian (positive definite metric) geometry \cite{Uhlenbeck}, was celebrated in the 2019 Abel Prize and 2007 Steele Prize.}  

Our extension of Uhlenbeck compactness states that any uniformly bounded sequence of connections in $L^\infty$ with curvature in $L^p$, has a convergent subsequence, strongly in $L^p$, weakly in $W^{1,p}$, under coordinate transformation to optimal regularity.  This applies both to sequences of affine connections $\Gamma_i$, as well as sequences of connections $(\Gamma_i,\A_i)$ defined on vector bundles $\mathcal{V}\mathcal{M}$ over $\mathcal{M}$, allowing for both compact and non-compact Lie Groups.  (See Theorem \ref{Thm_uhl} below.)   Uhlenbeck compactness provides a new starting point for analyzing the Einstein equations and Yang-Mills equations in Lorentzian geometry, at low regularity.  To give context, note that this can be viewed as a ``geometric'' improvement of the generalized Div-Curl lemma, which states that wedge products are weakly continuous when derivative bounds are given by exterior derivative, \cite{RoRoTemp}.   So if the components of a sequence of affine connections $\Gamma_i$ satisfy the constraint $\Gamma_i,d\Gamma_i$ uniformly bounded in $L^p$ in a given coordinate system, the Banach-Alaoglu Theorem  implies a subsequence of the connection components converges weakly in $L^p$, and the generalized Div-Curl lemma then implies that the components of curvature ${\rm Riem}(\Gamma_i)=d\Gamma_i+\Gamma_i\wedge\Gamma_i$ are weakly continuous on this limit, in $L^p$.   Our extension of Uhlenbeck compactness assumes, in addition, ``geometry'' in the form of the transformation law for connections, and  implies that, in addition, these connection components lie in $W^{1,p}$, (i.e., have one more derivative), and a subsequence converges {\it strongly} in $L^p$, weakly in $W^{1,p}$, but in a {\it different} coordinate system.   It is difficult to overstate the importance of strong over weak convergence in mathematical analysis--compactness theorems giving strong convergence are the essential starting point for establishing the validity of approximation schemes in nonlinear problems \cite{ChenSlemrod,Diperna,Tartar}.

Our program to address optimal regularity in this current general setting began with the insight that the metric singularities at GR shock waves might be due simply to the fact that the Riemann curvature tensor is formed from second derivatives of the metric tensor, but transforms like a first derivative object.   In fact, the problem lies more fundamentally at the level of an affine {\it connection}, the fundamental object more general than a metric, to which Riemann's theory of curvature applies.   The problem is this: because the transformation law for connections involves derivatives of the Jacobian, a transformation whose Jacobian has the same regularity as the connection, will in general transform a connection of {\it optimal} regularity, (one which has a level of regularity one derivative above its curvature tensor),  to a connection at the same regularity as its curvature, because the curvature maintains its regularity under tensor transformation by contraction with the undifferentiated Jacobian.\footnote{Note that the essential regularity of a connection, including whether it is optimal or non-optimal, is a geometric, coordinate independent notion when the manifold is restricted to the atlas of smooth coordinate transformations, but optimal regularity becomes coordinate dependent when the smooth atlas is extended to include low regularity coordinate transformations, c.f. \cite{ReintjesTemple_ell3}.}   Our conjecture, then, was that this was the {\it only} way non-optimal connections came into existence in geometry, including the singular metrics at GR shock waves.  But to prove the reverse direction, that non-optimal connections could always be smoothed to optimal regularity by coordinate transformation, one needs to undo the above process, and this requires constructing a suitable low regularity,  (\emph{singular} if you will), coordinate transformation, given only the information about the non-optimal connection and its curvature. For example, at the level of $L^{\infty}$ connections associated with shock waves, such a coordinate transformation $x\to y$ must be singular at shocks in the sense that jumps in derivatives of the Jacobian $J\equiv\partial y^{\alpha}/\partial x^i$ must be tuned to precisely cancel out the discontinuities in the given non-optimal connection in the transformation law for connections, i.e., we would need $J$ such that
\beq  \label{connection_transfo_TB}
 \Gamma^{\alpha}_{\beta\gamma}=\frac{\partial y^\alpha}{\partial x^i}\left\{\frac{\partial x^j}{\partial y^\beta}\frac{\partial x^k}{\partial y^\gamma}\Gamma^i_{jk}+\frac{\partial^2 x^i}{\partial y^{\beta}\partial{y^{\gamma}}}\right\},
\eeq
makes $\Gamma^{\alpha}_{\beta\gamma}$ Lipschitz continuous ($C^{0,1}$),\footnote{Our existence theory yields slightly weaker optimal regularity $\Gamma^{\alpha}_{\beta\gamma}\in W^{1,p}$, $p>n/2$, consistent with well known technical limitations of elliptic regularity theory in the space $L^\infty$, c.f. \cite{ReintjesTemple_Uhl1}.}  starting with $\Gamma^i_{jk}\in L^{\infty}.$  To see how this must work in the simpler case of metric connections, recall that by Christoffel's formula, metric components are always one derivative more regular than connection components in every coordinate system; so optimal metric regularity is two derivatives above the Riemann curvature, i.e., $g_{\alpha\beta}\in C^{1,1}$.     Thus the mapping $x\to y$ to optimal regularity would need to raise the regularity of components $g_{ij}\in C^{0,1}$ in $x$-coordinates, to $g_{\alpha\beta}\in C^{1,1}$ in $y$-coordinates, by the tensor transformation law,
\beq \nonumber
 g_{\alpha\beta}=\frac{\partial x^i}{\partial y^{\alpha}}\frac{\partial x^j}{\partial y^{\beta}}g_{ij}.
\eeq
This condition can only be met if the discontinuities in the derivatives $\frac{\partial}{\partial y^{\gamma}}g_{ij}$ on the right hand side are cancelled out by the derivatives of the Jacobians when the right hand side is expanded by the Leibniz product rule--a seemingly impossible condition to meet simultaneously for every $\gamma$ given the complexity of the discontinuity sets of arbitrary $L^\infty$ functions!   So what sort of equation would the Jacobians of such regularizing transformations satisfy if they exist at all?  And how would one find such an equation?  And what theory of mathematics would be available to solve such an equation?

The answer for general connections turned out to be the RT-equations, formulated to apply the now classical $L^p$ theory of elliptic regularity.
The RT-equations address the classical issue of non-optimal metrics and connections resulting from Riemann's formulation of curvature.   In General Relativity, the RT-equations reinforce Einstein's viewpoint of spacetime as a four dimensional geometry, in contrast to the $3+1$ approach required to apply classical methods of analysis in the study of the Cauchy problem.\footnote{Earlier investigations starting with Anderson \cite{Anderson}, (see \cite{ReintjesTemple_Uhl1}), addressed the problem of optimal regularity in GR from a $3+1$ point of view, based on using Kazdan and DeTurck \cite{DeTurckKazdan}  to regularize initial data on spacelike hypersurfaces where the gravitational metric restricts to positive definite, and then obtaining conditions (like the ``geodesic ball condition'') which control the regularity under time evolution.  Our work does not build on this.   Based on our stated conjecture here regarding the source of optimal regularity, our view from the start was that to get a theorem general enough to prove the optimal regularity of GR shock waves, we would have to build a new framework in which the connection and Riemann curvature tensor were treated fundamentally as $4$-dimensional geometrical objects.}   An announcement of our first results for affine connections focusing on the smooth case was published in \cite{ReintjesTemple_ell3}.    The theory of the RT-equations reached a culmination in our papers ``{\it On the regularity implied by the assumptions of geometry}'', \cite{ReintjesTemple_Uhl1,ReintjesTemple_Uhl2}, a title which conjures up Riemann's original habilitation.                         

\section{Singularities of spacetime}  It has been a central problem since Schwarzschild discovered his solution in 1915, whether a spacetime singularity in General Relativity is geometric, or whether it is removable by a regularizing coordinate transformation.   It took several years until Eddington proved that the apparent singularity at the Schwarzschild radius in Schwarzschild's solution, what we now call the event horizon of a black hole, was only a coordinate singularity.  At the time this question as to the nature of a spacetime singularity was essential in determining whether the equations of General Relativity, $G=8\pi T$, were to be taken seriously as a physically meaningful theory.  The celebrated Hawking-Penrose singularity theorems\footnote{Including the work for which Roger Penrose was awarded the 2020 Nobel prize in Physics \cite{Penrose}.} address the opposite side of this issue, providing conditions under which a spacetime singularity is {\it non-removable}.\footnote{Interestingly, Penrose's proof in \cite{Penrose} of the existence of a geodesic singularity requires the {\it apriori} assumption that the metric remain in $C^{1,1}$, which places the curvature in $L^\infty$, c.f. \cite{GrafGrant}.  Thus our theorems on optimal regularity would apply if the singularity formation actually involved a loss of optimal regularity, e.g., metric regularity falls to $C^{0,1}$ at the singularity, with curvature still in $L^\infty$.}    Our theory of the RT-equations in \cite{ReintjesTemple_Uhl1} resolves in the affirmative the open problem as to whether spacetime singularities are removable at the low regularity of $L^p$ curvature, thereby resolving the open problem as to the essential regularity of GR shock waves.  This theory is unrestricted by dimension or symmetry assumptions.  We note that removable black hole singularities, like the singularity at the Schwarzschild radius, are non-optimal solutions to which the RT-equations formally apply, but, unlike GR shock wave singularities,  black hole singularities lie below the threshold $L^p$ curvature assumptions in our current existence theory.

\section{Singularities at GR shock waves} In the year 1965 Glimm introduced the celebrated Glimm scheme of shock wave theory\footnote{Glimm's method was off-the-wall original and in stark contrast to the established method of analyzing hyperbolic PDE's at that time, {\it energy methods}. Even up until now, no one has succeeded in establishing Glimm's theorem by energy methods, or any method, even in one space dimension.}, and a year later Israel introduced the theory of Junction Conditions, the conditions under which two smooth gravitational metrics match Lipschitz continuously across an interface to form a general relativistic shock wave.  To solve the Einstein equations $G=8\pi T$ at a shock wave, the Einstein curvature tensor $G$ will inherit the regularity of the fluid which comprises the energy momentum tensor $T$, so since the density, pressure and velocity are discontinuous at shock waves, this places a discontinuity in the curvature of spacetime at the shock, as well.   The discontinuity in curvature is not a problem, but in Israel's theory,  the regularity of the gravitational metric was only one order above the curvature at shock surfaces constructed by the Junction Conditions, and this left open the possibility of so called ``delta function sources'' {\it in the second derivatives of the metric} at the shock.  Israel showed that ruling out the delta function sources in the Riemann curvature tensor at the shock was sufficient to imply the Rankine-Hugoniot jump conditions which enforce conservation at the shock, and under this assumption he proved that a low regularity transformation to Gaussian normal coordinates would regularize the apparent singularity in the metric and its connection at smooth shock surfaces, (see \cite{SmollerTemple} for a refinement of this result).  But the map to Gaussian normal coordinates is ill-defined as a continuous transformation when the shock surface is not smooth, i.e.,  when the normal vector is discontinuous, \cite{Reintjes}.   Israel's theory left open the problem of whether such singularities could be regularized at more complicated regions of interacting shock waves.   If not, then shock waves could create a new kind of singularity in GR, referred to by us as {\it Regularity Singularities} \cite{ReintjesTemple2}.

But after Israel there were no examples of such interacting shock waves in GR until 2005 when Groah and Temple first proved that the Glimm scheme could be extended to General Relativity in spherically symmetric spacetimes \cite{GroahTemple}.  This introduced into GR the first rigorous existence theory for shock waves admitting interactions of arbitrary complexity.  The validity of Glimm's method in GR re-introduced into General Relativity Israel's original problem with the Junction Conditions: The gravitational metric is singular at shock waves in the coordinate systems in which convergence of the Glimm scheme could be proven. The metric is singular in the sense that the connection, (defined in GR by the Christoffel symbols of the metric), degenerated to the level of regularity of its own Riemann curvature tensor, (discontinuous at shocks according to $G=8\pi T$), making the gravitational metric only one derivative more regular than its curvature. This level of regularity is so low that the existence of locally inertial frames, geodesic curves and the Newtonian limit, were at issue.   So as with Schwarzschild's original work, this raised the question as to whether these new shock wave solutions constructed by the Glimm scheme were physical.   It is well known that shock waves form generically when a fluid is sufficiently compressive, so the physical interpretation of these solutions addressed the basic physical status of the Einstein-Euler system.    This singular structure at shock waves either required a new physical interpretation for the Einstein-Euler equations, or else there was a missing theory of regularizing  coordinate transformations in General Relativity--some generalization of Israel's result to the case of complex shock wave interactions.  Interestingly, no one knows how to implement Glimm's method, or any other method for constructing shock wave solutions of the Einstein equations, in a coordinate system in which the shock waves exhibit optimal regularity.  

The fundamental question was then whether these solutions constructed by the Glimm scheme could be regularized by coordinate transformation at points of shock wave interaction.  I.e., does there exist some unknown theory of low regularity coordinate transformations sufficient to regularize these singularities?   To begin, Reintjes proved in \cite{Reintjes} that metrics could always be smoothed to optimal regularity at points of regular shock interaction in spherically symmetric spacetimes.   This was a complicated technical argument based on a non-local PDE, in which the Rankine-Hugoniot jump conditions came in again and again to make a system of seemingly over-determined PDE's, just barely solvable.  But the argument was highly tuned to the structure of the particular interaction, and gave no hints as to what, if any, was the general principle working behind the scenes \cite{Reintjes,ReintjesTemple1}.   

Our existence theory for RT-equations resolves this problem as follows: The extension of the Glimm scheme  in \cite{GroahTemple} establishes (weak) shock wave solutions of the Einstein-Euler equations in spherical symmetry. The Einstein-Euler equations, 
\beq \label{Einstein-Euler}
{\rm Div}_g(T)=0  \hspace{1cm} \text{and} \hspace{1cm}   G[g] = 8\pi T,
\eeq
couple the gravitational metric $g$ to the fluid variables (pressure $p$, density $\rho$ and velocity $u$) through the energy momentum tensor $T=(\rho+p)u\otimes u+pg$. Here ${\rm Div}_g$ denotes the covariant divergence of $g$ and $G[g]$ is the Einstein tensor, a $(0,2)$-tensor constructed from contractions of the Riemann curvature tensor. (By the Bianchi identities, $G=\kappa T$ imply ${\rm Div}(T)=0$, as a condition on the matter fields comprising $T$.)   The Einstein-Euler equations take a form simple enough to implement the Glimm method when the metric ansatz is taken to be Standard Schwarzschild Coordinates (SSC), 
\begin{eqnarray}\label{SSC}
ds^2=-B(r,t)dt^2+\frac{dr^2}{A(r,t)}+r^2d\Omega^2.
\end{eqnarray}
But the Glimm method only establishes the non-optimal metric regularity $A$ and $B$ Lipschitz continuous functions of $(r,t)$, and $L^{\infty}$ regularity in the Riemann curvature and fluid variables $\rho,p,u$.\footnote{The SSC coordinate system has played an important role in the history of GR, starting with Schwarzschild and Birkhoff.  Our theorem clarifies the logical status of the SSC coordinate system within GR. Indeed, solutions of the Einstein equations constructed in SSC,  are non-optimal at every order of regularity.  Non-optimality is built into SSC at the start by the choice of radial coordinate,  i.e., chosen to be proportional to the area of the sphere of symmetry.  Interestingly, the non-optimality of the SSC coordinate system has the effect of converting second order Einstein equations into first order equations.  Our theorem demonstrates that non-optimal metrics in SSC can be converted to optimal regularity by coordinate transformation at all levels of regularity, $L^p$ and above.  So on the one hand SSC is a degenerate non-optimal coordinate system, but intriguingly, it is precisely this degeneracy that makes it feasible to implement the first order Glimm scheme in SSC.}     

A natural setting which encompasses Israel's theory of junction conditions together with solutions constructed by Glimm's method is the setting of weak solutions of $G=\kappa T$ with $T\in L^{\infty}$, $g\in C^{0,1}$, and $\Gamma, {\rm Riem}(\Gamma) \in L^{\infty}$, c.f. \cite{Israel,GroahTemple}.  Our theorem establishes optimal regularity in a multi-dimensional setting general enough to include both cases.  Namely, our existence theorem for the RT-equations establishes that for any $\Gamma\in L^{\infty}$ with ${\rm Riem}(\Gamma)\in L^{\infty}$, there always exist coordinate transformations $x\to y$ which lifts the connection to $\Gamma\in W^{1,p}$, any $p<\infty$, and there is no need to assume the metric solves the Einstein equations.   The result thus establishes that in $y$-coordinates, the gravitational metric has optimal regularity  $g\in W^{2,p}$, any $p<\infty$.   As a corollary it follows that in transformed $y$-coordinates, a Lipschitz continuous solution of the Einstein equations would satisfy $G=\kappa T$ in the {\it strong} classical sense, geodesic curves do exist and locally inertial coordinates can be constructed. This  proves that such spacetimes are always {\it nonsingular}.   Interestingly, in contrast to SSC metrics of form (\ref{SSC}),  the Einstein equations for metrics in coordinates of optimal regularity appear too complicated to implement the Glimm scheme.

\section{Discovery of the RT-equations}   After entertaining in \cite{ReintjesTemple2} the possibility that there was some sort of new physics involved, authors became convinced that there must exist a fundamental, unknown theory of regularizing coordinate transformations, and we changed directions, and set out to find it.   This was the beginning of a decade long journey to discover the theory of these regularizing coordinate transformations.  

There was a controversy at the beginning and our program was viewed ``not feasible''.\footnote{This was the unanimous opinion by panels of experts at NSF Applied Mathematics.}  But fair enough.  After all, this program was proposing something completely new to General Relativity and Geometry:   The possibility that there might exist an as yet undiscovered {\it elliptic} system of equations which regularized the singularities in solutions of the {\it hyperbolic} Einstein-Euler equations.  We discovered these elliptic equations in \cite{ReintjesTemple_ell1}, and named them the ``Regularity Transformation equations'', or  RT-equations.\footnote{These equations were referred to as the Reintjes-Temple equations in \cite{ReintjesTemple_ell3}.}  The challenge then was to obtain a proof of existence of solutions to the RT-equations for the low regularity of GR shock waves, the space of connections in $L^\infty$ with $L^\infty$ Riemann curvature tensor. This space automatically extends Israel's condition that there be ``no delta functions sources'' at shocks to the case of general interacting shock waves.   Indeed, since Einstein built his equations to satisfy ${\rm Div}_g( T)=0$  by the Bianchi identities of geometry,  the condition that the curvature be locally bounded in $L^\infty$ naturally replaces the Rankine-Hugoniot jump conditions and the extension of this by the weak formulation of solutions in terms of test functions, as the condition on weak solutions which imposes conservation at shock waves in General Relativity, (see \cite{SmollerTemple}).   Our current existence theory for the RT-equations applies more generally to curvature in $L^p$, with connection in $L^{2p}$, $p>2,$ (or $p>n/2$ for $n$-dimensional manifolds).

Our first breakthrough for the general problem came in \cite{ReintjesTemple_geo}, where we obtained a condition on a non-optimal connection equivalent to the existence of a regularizing coordinate transformation.   We named this the {\it Riemann-flat condition}.   The Riemann-flat condition states that a regularizing coordinate transformation exists for a given non-optimal connection if and only if there exists a (1,2)-tensor $\tilde{\Gamma}$, one order more regular than the given non-optimal connection $\Gamma$, such that the connection $\Gamma-\tilde{\Gamma}$ is Riemann-flat, i.e.,      
\beq 
{\rm Riem}(\Gamma-\tilde{\Gamma})=0.     
\eeq  
Since this new Riemann-flat connection had the same shock structure as the original connection, we wondered for a while whether we might obtain optimal regularity by some sort of Nash embedding theorem.   The breakthrough came after we decided to try to construct an elliptic system of equations by  coupling equations for the unknown Jacobians  $J$, to equations for the above mentioned unknown (1,2)-tensor $\tilde{\Gamma}$, via the Riemann-flat condition and the coordinate Laplacian.  This opened the door to the discovery of the {\it Regularity Transformation} equations.

The new idea required to derive the RT-equations was to use the coordinate Euclidean metric instead of an invariant metric\footnote{The method of Kazden and DeTurck \cite{DeTurckKazdan} is based applying elliptic regularity to the operator which appears in the leading order part of the Ricci tensor as a consequence of the existence of an underlying invariant positive definite metric.  For non-Riemannian metrics this method leads to a wave type (hyperbolic) operator, and for general affine connections, there would be no associated invariant metric or elliptic operator.}  to define a co-derivative $\delta$, the idea being that the leading order part $d\Gamma$ of the Riemann curvature tensor in terms of a connection $\Gamma$, (the exterior derivative $d$ is independent of metric), can be converted into the elliptic Euclidean Laplacian via $\Delta=d\delta+\delta d$, where $\Delta = \partial^2_{x^1}+...+\partial^2_{x^n}$.   We then employ two equivalent forms of the Riemann-flat condition, one involving $d\tilde{\Gamma}$ and one involving $dJ,$ where $\tilde{\Gamma}$ is the unknown (1,2)-tensor which completes a non-optimal connection to Riemann-flat, and $J$ is the unknown Jacobian of the sought after regularizing coordinate transformation.  That is, by manipulating the Riemann-flat condition, we are able to show that equations for both $\tilde{\Gamma}$ and $J$ can be completed to form a pair of coupled nonlinear Poisson equations with left hand sides given by $\Delta\tilde{\Gamma}$ and $\Delta J,$ where $\Delta$ is the coordinate Laplacian constructed using the coordinate co-derivative $\delta.$   The RT-equations consist of these coupled Poisson equations, together with Cauchy-Riemann type equations in an auxiliary variable $A$ required to guarantee the integrability of $J$: 

\begin{align} 
\Delta \Gammati &= \delta d\Gamma - \delta \big(d J^{-1} \wedge dJ \big) + d(J^{-1} A ), \label{PDE1} \\
\Delta J &= \delta ( J \mm \Gamma ) - \langle d J ; \tilde{\Gamma}\rangle - A , \label{PDE2} \\
d \overrightarrow{A} &= \overrightarrow{\text{div}} \big(dJ \wedge \Gamma\big) + \overrightarrow{\text{div}} \big( J\, d\Gamma\big) - d\big(\overrightarrow{\langle d J ; \tilde{\Gamma}\rangle }\big),   \label{PDE3}\\
\delta \overrightarrow{A} &= v.  \label{PDE4}
\end{align}

The unknowns $(\Gammati,J,A)$ in the RT-equations, together with the given connection components $\Gamma$, are viewed as matrix valued differential forms defined by their components in some neighborhood $\Omega$ of a given coordinate system. Arrows denote ``vectorization'', mapping matrix valued $0$-forms to vector valued $0$-forms, (e.g. $\overrightarrow{A}^\mu = A^\mu_i dx^i$) and $\overrightarrow{\text{div}}$ is a divergence operation which maps matrix valued $k$-forms to vector valued $k$-forms.   The vector $v$ is free to impose and represents a new type of gauge freedom in the equations.  The operations on the right hand side are formulated in terms of the Cartan Algebra of matrix valued differential forms based on the Euclidean coordinate metric,  and the boundary condition $d\overrightarrow{J} =0$ must be imposed on $\partial\Omega$ for the auxiliary field $A$ to guarantee solutions satisfy $d\overrightarrow{J}=0$ in $\Omega$, (where $d\overrightarrow{J}\equiv {\rm Curl}(J)$), the vanishing curl condition implying that $J$ is integrable to coordinates, c.f. \cite{ReintjesTemple_ell1,ReintjesTemple_Uhl1}. 

To get the essence of how the RT-equations work, explained more carefully in Section \ref{Sec_Howitworks}, note first that smooth transformations preserve the regularity of connections, and their smooth Jacobians meet the RT-equations only by forcing $A$ to have the same regularity as the first two terms on the right hand side of  \eqref{PDE2}, in order to cancel out the irregularities and make the right hand side of \eqref{PDE2} also smooth.  To obtain, in contrast, coordinate transformations which lift the regularity of $\Gamma$ by one derivative, we apply elliptic regularity theory to prove the existence of solutions $A$ of \eqref{PDE3} - \eqref{PDE4} which are one order more regular than the first two terms on the right hand side of \eqref{PDE2}. This has the effect of lowering the regularity of $J$ and raising the regularity of $\Gammati$, to one derivative above the original connection $\Gamma$, and from this we ultimately deduce optimal regularity of the transformed connection $\Gamma$. This is how the RT-equations, formally derivable for smooth coordinate transformations, produce low regularity transformations which lift connections to optimal regularity, i.e., to one derivative above their curvature.

The formulation of the RT-equations (\ref{PDE1})-(\ref{PDE4}), the case of affine connections, was directed toward making them amenable to the classical $L^p$ methods of elliptic regularity theory, a regularity low enough to include GR shock waves constructed by Glimm's method in \cite{GroahTemple}, solutions of the Einstein-Euler equations with fluid density, velocity, and Riemann curvature in $L^\infty$.   Our publication in \cite{ReintjesTemple_ell1} is devoted to a careful derivation of the RT-equations, and in \cite{ReintjesTemple_ell2} we establish their viability by proving the first existence theory for the RT-equations applicable to non-optimal connections in $W^{m,p}$, $m\geq1$, a level of regularity $m$ orders above the regularity of GR shock waves.  In order to extend this existence theory to the low regularity of $L^p$, we use the gauge-type freedom associated with $v$ in (\ref{PDE4}) to identify the role played by gauge-type transformations associated with a change of $v$ in the equations.   By a serendipitous choice of $v$, we construct a solution dependent gauge transformation which de-couples equations (\ref{PDE2})-(\ref{PDE4}) from (\ref{PDE1}), to form a subsystem which we call the {\it reduced} RT-equations,  a system of elliptic PDE's amenable to an existence theory at a level of regularity one order below what we could achieve with the original RT-equations. By this we succeed in extending our existence theory to $L^{2p}$ connections with $L^p$ curvature, $p>n/2$.   This level of regularity is low enough to include both GR shock waves constructed by the Glimm scheme, as well as shock solutions constructed in multi-d by Israel's theory of junction conditions, i.e., all of the known examples of GR shock waves \cite{ReintjesTemple_ell3}.             

In \cite{ReintjesTemple_Uhl1}, we extend the theory of the RT-equations from the setting of affine connections on the tangent bundle of a manifold ${\mathcal M},$ to connections  $(\Gamma,\A)$ on vector bundles ${\mathcal V}{\mathcal M}$ over a base manifold $\mathcal M$, where $\A$ denotes the connection on the fibre ${\mathbb R}^N$, and $\Gamma$ is a non-optimal affine connections  on the base manifold ${\mathcal M}$.  Solutions of the vector bundle version of the RT-equations produce gauge transformations which lift the regularity of non-optimal connections on the fibre, up to optimal regularity.  The RT-equations only involve the uncoupled leading order part $(d\A,d\Gamma)$ of the curvature $2$-form. As a result the RT-equations decouple in $\Gamma$ and $\A$, even though they remain coupled through the  lower order commutator term in the curvature. The vector bundle version of the RT-equations is given by
\begin{eqnarray}
\Delta \Ati &=& \delta d \A -\delta\big( dU^{-1}\wedge dU \big),   \label{RT_1_intro} \\
\Delta U &=&   U \delta\A   -  (U^T \eta )^{-1}  \langle dU^T ; \eta dU\rangle . \label{RT_2_intro}
\end{eqnarray}
Here $U$ is the sought after regularizing gauge transformation, $\Ati$ is an auxiliary matrix valued $1$-form from which the connection of optimal regularity can be derived, and $\eta$ is the matrix which specifies the gauge group.  

\section{Optimal Regularity}   Our existence theory for the RT-equations (\ref{PDE1})-(\ref{PDE4}) and (\ref{RT_1_intro})-(\ref{RT_2_intro}) is a new application of the now classical $L^p$ theory of elliptic regularity, a theory set out by the great analysts of the 1950's and 60's, including Agmon, Nirenberg, Lax, Milgram and others, \cite{GilbargTrudinger}.   Our existence theory for \eqref{RT_1_intro} - \eqref{RT_2_intro} in \cite{ReintjesTemple_Uhl2} establishes the existence of gauge transformations which lift connections $\A \in L^{2p}$ with $d\A\in L^p$, $p>n/2$, to optimal regularity $\A\in W^{1,p}$, for connections $\A$ on the fibre of a vector bundle.  This argument is analogous to the (more complicated) theory of (\ref{PDE1})-(\ref{PDE4}) which establishes the existence of affine connections of optimal regularity on the tangent bundle of a manifold, \cite{ReintjesTemple_Uhl1}.\footnote{The results in \cite{ReintjesTemple_Uhl1,ReintjesTemple_Uhl2}  carry over to every order of regularity at or above $L^p$, c.f. \cite{ReintjesTemple_ell2}.}     The existence theory in \cite{ReintjesTemple_Uhl2} for equations (\ref{RT_1_intro})-(\ref{RT_2_intro}), taken together with the existence theory in \cite{ReintjesTemple_Uhl1} for  (\ref{PDE1})-(\ref{PDE4}), proves optimal regularity for general non-optimal connections $(\Gamma,\A)$ on vector bundles $\mathcal V\mathcal M.$    The argument applies to both compact and non-compact gauge groups, in particular $SO(r,s)$ and $SU(r,s),$ \cite{ReintjesTemple_Uhl2}.   For concreteness, we restrict to $SO(r,s),$ $r+s=N\geq0,$ the group of special orthogonal matrices with respect to a metric $\eta$  of signature $(r,s)$, a group which is non-compact for $r,s>0$, and compact when $r=0$ or $s=0.$  The following theorem is a synthesis of Theorem 2.1 in \cite{ReintjesTemple_Uhl1} and Theorem 2.3 in \cite{ReintjesTemple_Uhl2}.

\begin{Thm} \label{Thm_opt}  
Let ${\mathcal V\mathcal M}$ be a vector bundle over an $n$-dimensional manifold ${\mathcal M}$ with $N$-dimensional fibre acted on by a compact or non-compact Lie Group (take $SO(r,s)\subset {\mathbb R}^{N\times N}$), let $(x,\textbf{a}( x))$  be a given coordinate system $x$ on ${\mathcal M}$  paired with a given gauge $\textbf{a}$, (a pointwise basis on the fibres ${\mathbb R}^N$),  and assume connection $(\Gamma,\A)$ on ${\mathcal V\mathcal M}$ satisfies
\beq \label{bound_1}
\| (\Gamma_x,\A_{(x,\textbf{a})}) \|_{L^{2p}} + \|(d\Gamma_x,d\A_{(x,\textbf{a})}) \|_{L^p} \; \leq \; M,
\eeq
where norms are taken components-wise in $x$-coordinates, $p>\max\{n/2,2\}$, $n \geq 2$, $p<\infty$. Then, locally, there exists a coordinate transformation $x \to y$ with Jacobian $J=\frac{\partial y}{\partial x}$ and a gauge transformation $U \in SO(r,s)$, such that,  in $y$-coordinates, the connection components $(\Gamma_y,\A_{(y,\textbf{b})})$  in gauge $\textbf{b} = U \cdot \textbf{a}$ exhibit optimal regularity $(\Gamma_y,\A_{(y,\textbf{b})}) \in W^{1,p}$, and satisfy
\beq \label{bound_2}
\|(\Gamma_y,\A_{(y,\textbf{b})}) \|_{W^{1,p}} 
\; \leq\; C(M) ,
\eeq
where $C(M) > 0$ is a constant depending only on $M$.   The transformations $J$ and $U$ accomplishing this satisfy\footnote{The regularity of $J$ and $U$ may be measured in $x$- or $y$-coordinates.}
\begin{align} \label{bound_3}
\|J\|_{W^{1,2p}}  + \|J^{-1}\|_{W^{1,2p}} + \|U\|_{W^{1,2p}}  + \|U^{-1}\|_{W^{1,2p}} 
\; \leq\; C(M) ,
\end{align}
where $C(M)$ can be taken to be the same constant.         
\end{Thm}

\section{Uhlenbeck compactness}     As a serendipitous corollary of Theorem \ref{Thm_opt}, the extra derivative implied by (\ref{bound_2}) of Theorem \ref{Thm_opt} directly implies Uhlenbeck compactness for both compact and non-compact gauge groups acting on the fibre, incorporating general affine connections and Lorentzian metrics on the base manifold.   The following theorem synthesizes the results in \cite{ReintjesTemple_Uhl1} and \cite{ReintjesTemple_Uhl2}.  It gives the most general statement of Uhlenbeck compactness for vector bundles at the lowest level of $L^p$ curvature, implied by Theorem 2.2 in \cite{ReintjesTemple_Uhl1} and Theorem 2.4 in  \cite{ReintjesTemple_Uhl2}. 

\begin{Thm}\label{Thm_uhl}
Let $(\Gamma_i,\A_i) \in L^\infty$ be a sequence of connections on ${\mathcal V}\mathcal{M}$ uniformly bounded by
\beq \label{bound_4}
\| (\Gamma_i,\A_i) \|_{L^{\infty}} + \|(d\Gamma_i,d\A_i) \|_{L^p} \; \leq \; M
\eeq
in a given coordinate system and gauge $(x,\textbf{a})$,  $p>n$, (c.f. \eqref{bound_1}). Then, under coordinate and $SO(r,s)$-gauge transformation to optimal regularity, a subsequence of the transformed connections $(\Gamma_i,\A_i)$ converges strongly in $L^p$, weakly in $W^{1,p},$ to a limit connection $(\Gamma,\A)$.\footnote{More precisely,  after extraction of a subsequence, the $x\to y_i$ coordinate transformations to optimal regularity at each $i\in{\mathbb N}$ converge $y_i\to y$ weakly in $W^{2,2p}$, (c.f. (\ref{bound_3})), and the corresponding $y_i$ connection components converge $(\Gamma_i,\A_i)\to (\Gamma,\A)$ weakly in $W^{1,p}$, (c.f. (\ref{bound_2})), when expressed as functions of the original $x$-coordinates.}  
\end{Thm}

Theorem \ref{Thm_uhl} extends Uhlenbeck compactness from Uhlenbeck's setting of connections on vector bundles over Riemannian manifolds with compact gauge groups, to connections on vector bundles over general base manifold, allowing for both compact and non-compact gauge groups, including the Lorentzian geometries of Relativistic Physics.   Taken together with our results in \cite{ReintjesTemple_ell2,ReintjesTemple_ell3}, this establishes Uhlenbeck compactness at every order of regularity at or above $L^p$.  As a first example for how one might apply Uhlenbeck compactness in General Relativity, we give in \cite{ReintjesTemple_Uhl1} a new compactness theorem for approximate solutions of the Einstein equations in vacuum spacetimes, establishing convergence to a solution of the Einstein equations by using that the strong $L^p$ convergence asserted by Theorem \ref{Thm_uhl} is sufficient to pass limits through nonlinear products.  We expect this principle to extend to non-vacuum solutions when combined with further analysis of the matter field equations, for example the Einstein-Euler equations, by combining Uhlenbeck compactness with the method of compensated compactness.

Uhlenbeck compactness is ideal for application to geometric PDE's, including the relativistic Einstein equations and Yang-Mills equations, because compactness follows from estimates based on the curvature alone, the observable in the theory, without having to deal with the other potentially uncontrolled derivatives of the connection; and the regularity of convergence obtained is precisely what is needed to pass limits through nonlinear products.  Compactness is the starting point of analysis, and there are scant few fundamental compactness theorems applicable to hyperbolic PDE's.  This new Uhlenbeck compactness result offers a fresh starting point for the analysis of the Einstein and Yang-Mills equations in Lorentzian geometry.

\section{How the RT-equations establish Optimal Regularity}  \label{Sec_Howitworks}

We now explain how solutions of the RT-equation furnish the coordinate and gauge transformations to optimal regularity. Detailed proofs can be found in \cite{ReintjesTemple_Uhl1,ReintjesTemple_Uhl2}. Since the connection components $(\Gamma,\A)$ act independently on tangent and vector bundles separately, and are uncoupled in the RT-equations \eqref{PDE1} - \eqref{PDE4} and \eqref{RT_1_intro} - \eqref{RT_2_intro}, it suffices to establish optimal regularity for $\Gamma$ and $\A$ separately.   

We first address the case of vector bundles, c.f. \cite{ReintjesTemple_Uhl2}. So assume a non-optimal connection $\A_\textbf{a} \in L^{2p}(\Omega)$ with $d\A_\textbf{a} \in L^p(\Omega)$ in gauge $\textbf{a}$. Our existence theory in \cite{ReintjesTemple_Uhl2} yields solutions $U \in W^{1,2p}(\Omega)$ of the so-called \emph{reduced} RT-equation \eqref{RT_2_intro}, on potentially restricted domains $\Omega$, by an iteration scheme approximating \eqref{RT_2_intro} by Poisson equations with Dirichlet data in $SO(r,s)$. The condition that any such solution $U$ is indeed a gauge transformation in $SO(r,s)$, is a built-in property of the reduced RT-equation \eqref{RT_2_intro}. Namely, introducing $w \equiv U^T \eta U-\eta$, the RT-equations \eqref{RT_2_intro} in combination with the Lie Algebra condition $\delta \A_\textbf{a}^T \cdot \eta + \eta \cdot \delta \A_\textbf{a} =0$, imply that $w$ is a solution of
\beq \label{SO(n)_eqn2}
\Delta w = \delta \A_\textbf{a}^T \cdot w + w \cdot  \delta \A_\textbf{a}  
\eeq
with zero Dirichlet data. We then prove, in the presence of non-uniqueness due to a Fredholm alternative, that our iteration scheme generates only the trivial solution $w=0$ of \eqref{SO(n)_eqn2}, which by definition is equivalent to the desired gauge condition $U \in SO(r,s)$, c.f. \cite{ReintjesTemple_Uhl2}.

We now explain how solutions $U \in W^{1,2p}\big(\Omega,SO(r,s)\big)$ of the reduced RT-equation   \eqref{RT_2_intro} map a non-optimal connection $\A_\textbf{a} \in L^p(\Omega)$ to a connection of optimal regularity $\Aop \in W^{1,p}(\Omega)$ in gauge $\textbf{b} = U\cdot \textbf{a}$, under the connection transformation law
\beq \label{connection_transfo_VB}
\A_\textbf{a} = U^{-1} dU + U^{-1} \Aop U.
\eeq
For this we show in \cite{ReintjesTemple_Uhl2} that $\Ati' \equiv \A_\textbf{a} - U^{-1} dU$ defines a solution of the first RT-equation \eqref{RT_1_intro},
\begin{eqnarray}\label{RT_Gammati'}
\Delta \Ati' &=&  \delta d \A_\textbf{a} - \delta (dU^{-1} \wedge dU) ,
\end{eqnarray}
by substitution of the right hand side of the reduced RT-equation \eqref{RT_2_intro} for $\Delta U$. This substitution results in a cancellation in uncontrolled terms involving $\delta \A_\textbf{a}$, which then entails the regularity gain of one derivative to $\Ati' \in W^{1,2p}(\Omega)$ by application of interior elliptic estimates to \eqref{RT_Gammati'}. Introducing then 
\beq \label{OptimalGamma}
\Aop \equiv U \Ati' U^{-1} =  U \A_\textbf{a} U^{-1} -  dU \cdot U^{-1}, 
\eeq
we find by comparison with \eqref{connection_transfo_VB} that $\Aop$ is indeed the connection of optimal regularity in gauge $\textbf{b} = U\cdot \textbf{a}$.    
\vspace{.2cm}

To reiterate, solving the reduced RT-equations yields the regularizing transformation, and the first RT-equation then yields the gain of one derivative to optimal regularity.  This principle carries over to connections on tangent bundles. However, because the first RT-equation \eqref{PDE1} is coupled  through $\Gammati$ to the others \eqref{PDE2} - \eqref{PDE4}, it was a major obstacle for authors to identify an analogous system of reduced RT-equations applicable to the affine case \eqref{PDE1} - \eqref{PDE4}. In particular, at the start there was no indication that such a system even existed.  It was because our earlier iteration scheme in \cite{ReintjesTemple_ell2,ReintjesTemple_ell3} did not close at the low regularity of $L^p$ connections, due to non-linearities in the first RT-equation \eqref{PDE1}, that we eventually discovered internal ``gauge''-type transformations on the space of solutions of the RT-equations \eqref{PDE1} - \eqref{PDE4}, that enabled us to separate off \eqref{PDE1} from the remaining equations. This latter system is what we refer to as the \emph{reduced} RT-equations \cite{ReintjesTemple_Uhl1}, 
\begin{eqnarray} 
\Delta J &=& \delta ( J \mm \Gamma ) - B , \label{RT_withB_2} \\
d \vec{B} &=& \overrightarrow{\text{div}} \big(dJ \wedge \Gamma\big) + \overrightarrow{\text{div}} \big( J\, d\Gamma\big) ,   \label{RT_withB_3} \\
\delta \vec{B} &=& w.  \label{RT_withB_4}
\end{eqnarray}
The reduced RT-equations \eqref{RT_withB_2} - \eqref{RT_withB_4} are linear in $(J,B)$, and consequently our iteration scheme (based on Poisson equations) in \cite{ReintjesTemple_Uhl1} extends to the low regularity of $L^p$ connections, establishes existence of solutions $(J,B)$ in a neighborhood $\Omega$ of any point. It is a built-in property of \eqref{RT_withB_2} - \eqref{RT_withB_4} that any solution $J$ is a Jacobian integrable to coordinates, as long that the integrability condition $d\vec{J}\equiv {\rm Curl}(J)=0$ holds on the boundary $\partial\Omega$ (accomplished by our existence theory in \cite{ReintjesTemple_Uhl1}). That is, combining \eqref{RT_withB_2} with \eqref{RT_withB_3}, a computation shows that $w\equiv d\vec{J}$ is a solution of the Laplace equation $\Delta w=0$, which together with our boundary data implies that $w=0$ throughout $\Omega$. This implies that $J$ is a true Jacobian integrable to coordinates. 

Given now a solution $(J,B)$ of the reduced RT-equation \eqref{RT_withB_2} - \eqref{RT_withB_4} with integrable Jacobian $J$, one recovers a solution $(J,\Gammati,A)$ of the full RT-equations \eqref{PDE1} - \eqref{PDE4} by introducing
\beq  \label{Gammati'} 
\Gammati \equiv \Gamma - J^{-1} dJ, 
\hspace{.5cm} 
A \equiv B - \langle d J ; \Gammati \rangle ,  
\hspace{.5cm} \text{and} \hspace{.5cm}
v \equiv w - \delta \overrightarrow{\langle d J ; \Gammati \rangle},
\eeq
as can be verified by direct computation using \eqref{RT_withB_2} to eliminate uncontrolled terms involving $\delta\Gamma$. From interior elliptic estimates, applied to the first RT-equations \eqref{PDE1}, one then obtains that $\Gammati$ is in $W^{1,p}$, a gain of one derivative over $\Gamma$. Defining 
\beq \label{Gamma_y_reverse}
(\Gamma_y)^\gamma_{\alpha\beta} \equiv J_k^\gamma (J^{-1})^i_\alpha  (J^{-1})^j_\beta   \; \Gammati^k_{ij},
\eeq
a comparison of \eqref{Gammati'} with the connection transformation law \eqref{connection_transfo_TB} tells us that $\Gamma_y$ is indeed the transformed connection of optimal regularity, $\Gamma_y \in W^{1,p}(\Omega)$. These are the essential ideas underlying the proof of Theorem \ref{Thm_opt}. We refer the interested reader to \cite{ReintjesTemple_Uhl1} and \cite{ReintjesTemple_Uhl2} for a complete proof, worked out in full detail, starting with a suitable notion of weak solution.
\vspace{.2cm}

Finally we note that one can construct examples of non-optimal connections from connections of optimal regularity by applying low regularity coordinate or gauge transformations. The inverse transformation is consequently a transformation to optimal regularity and, by the derivation of the RT-equations, must be a solution of the RT-equations for some $v\in L^p$ in \eqref{PDE4}.\footnote{Note, in the case of the RT-equations on vector bundles, we set the analog of $v$---a Lie Algebra valued one form $X\in L^p$---to zero in \eqref{RT_1_intro} - \eqref{RT_2_intro}, as this is of no relevance for establishing optimal regularity by the RT-equations in Theorem \ref{Thm_opt}, c.f. \cite{ReintjesTemple_Uhl2}.} The interested reader may use this to verify the validity of the RT-equations and Theorem \ref{Thm_opt} in explicit examples, but we expect such examples do not provide any deeper insights beyond the more general derivation of the RT-equations in \cite{ReintjesTemple_ell1,ReintjesTemple_ell3} and \cite{ReintjesTemple_Uhl2}.

\section{Perspectives on Optimal Regularity in the context of the Cauchy Problem in GR}  

Current methods of the Cauchy problem appear to be inadequate for establishing optimal regularity and for explaining how non-optimal solutions fit into the general picture. For concreteness, consider Theorem 1.6 in \cite{Klainermann}, which provides $L^2$ estimates for solutions assumed to exist {\it apriori}.  This applies to vacuum solutions of the Einstein equations, an assumption which excludes shock waves, the primary interest of our papers.  However, what is interesting in light of our new result, is that Theorem 1.6 in \cite{Klainermann} does not apply in general to non-optimal solutions with $L^p$ curvature, $p\geq 2$, unless the second fundamental form is one derivative more regular than the curvature on a Cauchy Surface. But the second fundamental form has in general only the regularity of the spacetime connection, and when the connection is non-optimal, the second fundamental form lacks one derivative required for Theorem 1.6 in \cite{Klainermann} to apply. Said differently, the theory in \cite{Klainermann}, like other $3+1$ Cauchy problem based methods, treat non-optimal solutions as anomalies, and at the lowest regularity of $L^p$ curvature, give results only for solutions assumed to exhibit optimal regularity in spacetime at the start.\footnote{The intrinsic metric on the Cauchy surface is regularized in \cite{Klainermann}  at the start by the method of Kazdan-DeTurck \cite{DeTurckKazdan}, but this method does not in general regularize the second fundamental form on the Cauchy surface. Our theory shows that optimal regularity requires in general spacetime transformations (obtained by solving the RT-equations), to regularize a $4$-dimensional metric.}   

Referees of our papers raised the objections that non-optimal solutions could eventually be ruled out by {\it well-posedness} considerations, or that the initial value problem will automatically regularize all non-optimal solutions, or that non-optimal solutions are just an anomaly of spherical symmetry.  Based on the theory of optimal regularity derived from the RT-equations, these statements are {\it not correct}, (c.f. our discussions in  \cite{ReintjesTemple_ell2,ReintjesTemple_ell3}).     

To clarify, non-optimal solutions exist in spherically symmetric spacetimes for the same reason they exist in general spacetimes--the Riemann curvature tensor is a second derivative construct which transforms like a tensor. So start with the fact that non-optimal initial data and non-optimal solutions exist due to the tensorial nature of Riemann's curvature tensor.  Then it is easy to conjecture that all optimal solutions do not ``fit'' within any ``one'' coordinate ansatz, in the sense that there is no single coordinate {\it ansatz} which simultaneously lifts all non-optimal solutions to optimal regularity at once.   That is, you have to solve the RT-equations, an essentially $4$-dimensional, not $3+1$ dimensional system.  Without the RT-equations, the Cauchy problem in GR is thus faced with the problem as to what to do with non-optimal solutions: I.e., they aren't regular enough to evolve in time, but if they are ``physical'', you can't neglect them either.   Now at higher levels of regularity, say connection and curvature of regularity above $W^{1,p}$, one can simply accept estimating non-optimal solutions as one order less regular than they really are---the RT-equations could be used to regularize such solutions to optimal---, but at the lowest level of regularity, there is no room left, in the sense that $L^p$  connections with $L^p$ curvature are not regular enough to evolve in any regularity class.  

Now if one did not know, as our new theorem demonstrates (above $L^p$, $p>n/2$), that non-optimal solutions are always perfectly ``good'' optimal solutions written in a ``bad'' coordinate system, then it might make sense to rule out the bad one's by ``well-posedness'' considerations.   After all, if the connection essentially lacks that extra derivative of regularity, it won't be stable in derivative norms at that level.    The hope, then, would be the idea that some one gauge for the initial value problem would somehow make ``good'' solutions optimal, and rule out ``bad'' ones by ``well-posedness''.  

Unfortunately for this idea our new theorem tells us that {\it all} of the non-optimal solutions are actually ``good'' in some other coordinate system, (our current theorem applies to curvatures in $L^p$, $p>2$), but the coordinate transformation which regularizes a given non-optimal solution,  is highly tuned to that particular solution, (you have to solve the RT-equations), so there won't be any one {\it ansatz} that regularizes them ``all at once''.   On the other hand, by properties of the regularizing coordinate transformations obtained by solving the elliptic RT-equations,  one should expect two non-optimal solutions with connection and curvature ``close in the $L^p$ sense'', will be mapped to two solutions ``close'' in $W^{1,p}$.   So if one didn't know about the RT-equations,  one might well be puzzling for a long time over ruling out solutions by ``well-posedness'' considerations, when the issue is basically a problem of optimal regularity instead.   

Our conclusion, then,  is that our proof that non-optimal solutions in $L^p$ can always be regularized by low regularity coordinate transformations, fills a gap in the understanding of the role played by non-optimal solutions in current theories of the Cauchy problem.  

\section{Outlook on Numerical Relativity} Our results give an $L^p$ based answer to the question as to whether loss of optimal regularity under time evolution in numerical or theoretical approximations, (one example being the formation of shock waves), represents the formation of real spacetime singularities, or simply a loss of regularity in the coordinate system employed.  By this new result, solutions can always be regularized to optimal regularity by coordinate transformation in neighborhoods where the components of the connection and curvature are locally $L^p$ functions.   The iteration scheme in our existence theory, based on solving linearized Poisson equations,  provides an explicit numerical algorithm for constructing the Jacobians $J$ of the regularizing coordinate transformation, and the optimal connection is then given by an exact formula. We believe this theory for regularizing singularities should find applications in Numerical Relativity.      

\section*{Funding}

M. Reintjes has been partially supported by CityU Start-up Grant for New Faculty (7200748) and by CityU Strategic Research Grant (7005839).

\section*{Acknowledgement}

We are grateful to Craig Evans for pointing us to several important references on $L^p$ elliptic regularity theory.

\end{document}